\documentstyle[12pt,epsf]{article}
\begin{document}

\begin{center}
{\Large\bf NUCLEON FORM FACTORS IN 

A RELATIVISTIC THREE-QUARK MODEL}

\vspace*{1cm}
{\large\bf M.A. Ivanov, V.E. Lyubovitskij\\}

\vspace*{.2cm}
{\em Bogoliubov Laboratory of Theoretical Physics, \\
Joint Institute for Nuclear Research,\\
141980 Dubna (Moscow region), Russia\\}

\vspace*{.4cm}
{\large\bf M.P. Locher\\}

\vspace*{.2cm}
{\em Paul Scherrer Institute, CH-5232 Villigen PSI}
\end{center}

\abstract{
We report the calculation of electromagnetic form factors of
nucleons within a relativistic three-quark model with Gaussian shape for the
nucleon-quark vertex. The allowed region for two adjustable parameters,
the range parameter $\Lambda_N$ in the Gaussian and the constituent quark
mass $m_q$, is obtained from fitting the data
for magnetic moments and electromagnetic radii of nucleons.
It is found that their values calculated with $m_q$=420 MeV and
$\Lambda_N$=1.25 GeV agree very well with the experimental data.

It is turned out that the electric proton and magnetic nucleon form factors
fall faster than the dipole fit of the experimental data
for momentum transfers $0\le Q^2\le 1$ GeV$^2$.
}

\begin{center}
\end{center}

\section{Introduction}

Electromagnetic properties of nucleons are an important source of information
on the internal structure of baryons.
The success of the nonrelativistic quark model for
the description of static characteristics (masses, magnetic moments, etc.)
and the results from deeply inelastic lepton
scattering are a clear indication for the three-quark structure of nucleons.

In view of the difficulties of describing the nucleon as a relativistic three
quark system rigorously, on the basis of QCD, many methods and models have been
developed which implement important aspects of QCD at least partially. We
mention a few.
Some approaches \cite{Rober,AndNov} are formulated using
classical hadronic degrees of freedom (color singlets) without explicit
reference to quarks and gluons. This holds in
particular for chiral perturbation theory which addresses low energy scattering
of hadrons \cite{Gass}. In the Skyrmion model the
baryon is considered to be a solitonic excitation of classical meson fields
which are described by
an effective chiral Lagrangian, see \cite{Meis}, e.g. The
electromagnetic
radii and magnetic moments of nucleons as well as the behavior of
electromagnetic form factors up to 1 GeV$^2$ are described in this way.

In the context of quark models the nonrelativistic approximation is
problematic even for the constituent-quark picture at low energies
since the effective masses
and the intrinsic momenta have the same order of magnitude.
An attempt to implement relativistic invariance
for the description of
the electromagnetic properties of the nucleon is the
covariant constituent quark models by Konen and Weber \cite{Weber},
and Chung  and Coester \cite{Coester} which use light front dynamics
for the constituent quarks. In this framework the available
data for $0\le Q^2 \le 1.5$ GeV$^2$ \cite{Weber} and $0\le Q^2 \le 6$ GeV$^2$
\cite{Coester} have been well described with two adjustable parameters
(constituent quark mass and confinement scale).

For large momentum transfers, $Q^2>>m^2_N$, perturbative QCD predicts
that the magnetic form factor behaves as $G_M\sim 1/Q^4$ \cite{Brod}.
The use of QCD sum rules allows to extrapolate the electromagnetic
form factors to low and moderate $Q^2$ by incorporating
local quark-hadron duality \cite{Rad}.
Recall that the calculation of hadronic form factors by sum rules in their
original version \cite{Shif} is restricted to
intermediate momentum transfers
and does not work in the infrared region due to power corrections $1/Q^{2n}$.
Therefore, in order to calculate magnetic moments of nucleons
QCD sum rules in the external field approach have been introduced
\cite{IofSmil}. For a
general method of calculating the nucleon magnetic form factors at small
$Q^2$ see \cite{Belyaev}.

For applications at high energies the diquark model \cite{Kroll} has been
proposed. In this model the proton is built from quarks and diquarks and the
diquark is treated as a quasi-elementary particle. Fits to the data
for $Q^2>3$ GeV$^2$ \cite{Expt} with few parameters have been obtained.

A different line of attack uses both quark {\it and} hadronic degrees of
freedom.
Chiral potential models (see, for example, \cite{Baric} and references therein)
fall into this class. They use an effective confining potential
for the quarks and a quark pion interaction which
preserves chiral symmetry. Incorporating the lowest-order pionic
correction, the magnetic moments of the nucleon octet have been calculated
\cite{Baric}. Alternative approaches \cite{Goldman}-\cite{Efi2} start from
{\it effective Lagrangians} which
describe the transition of hadrons to their constituent quarks, combined
with some
assumptions on the behavior of quarks at low energies and on the
shape of the hadron-quark
form factor. This approach is connected fairly directly to the difficult
issues of hadronization and quark confinement in QCD.

In the present paper we extend the line of thought developed in \cite{FBS}
and \cite{Diquark} which use local hadron-quark or
local hadron-quark-diquark vertices together with a confined quark propagator
(no poles). It has been shown for these models that the electromagnetic
form factors of nucleons tend to be somewhat above the data for
$Q^2<1$ GeV$^2$. This might be a hint that the assumption of a
{\it local} coupling of baryons with their constituents is not quite
adequate. Introducing a {\it nonlocal}
hadron-quark vertex may be a good way of effectively introducing other
degreees of freedom (mesonic cloud, soft gluons, etc.). The Nambu-Jona-Lasinio
models, NJL, with separable interactions are candidates for this purpose.
Many successful applications of such  models for low-energy pion physics
exist (see, e.g., \cite{Goldman,Gross} and references therein).

In \cite{Aniv1,Aniv2} the Lagrangian formulation of the NJL model
with separable interaction has been given both for mesons and (for the first
time) for baryons. The gauging of nonlocal interactions has been done
by using the time-ordering P-exponent. The pion weak decay constant,
the two-photon decay width, as well as the form factor of the
$\gamma^*\pi^0\to\gamma$ transition, the pion charge form factor, and
the strong $\pi NN$ form factor have been calculated and good
agreement with the data has been achieved with three parameters,
the range parameters characterizing the size of mesons $\Lambda_M$ and
baryons $\Lambda_B$, and the constituent quark mass $m_q$. In this talk
we report the results for the electromagnetic form factors of
nucleons \cite{PSI} within the approach developed in \cite{Aniv1,Aniv2}.
First, we slightly modify the gauging of the nonlocal quark-hadron vertex by
using a path-independent definition for the derivative of the
time-ordering P-exponent \cite{Mand,Tern}. Second, we derive  Feynman rules
for the diagrams which describe the electromagnetic nucleon form factors.
For simplicity, we use a Gaussian shape for the nucleon-quark vertex.
The permissible range for the two adjustable parameters, the Gaussian range
$\Lambda$ of the separable interaction and the
constituent quark mass $m_q$, is obtained by fitting the experimental values
of the magnetic moments and the electromagnetic radii.
It is found that their values calculated with $m_q$=420 MeV and
$\Lambda_N$=1.25 GeV agree very well with the experimental data.

It is turned out that the electric proton and magnetic nucleon form factors
fall faster than the dipole fit of the experimental data
for momentum transfers $0\le Q^2\le 1$ GeV$^2$.

\section{Model}

We start with a brief review of our approach. We consider the hadron as being
a bound state of relativistic constituent quarks with masses $m_q$
\cite{Aniv1}-\cite{PSI}. The transition of hadrons into constituent quarks
and {\it vice versa} is
described by the corresponding interaction Lagrangian.
The interaction Lagrangian of nucleons with quarks is written as
\begin{eqnarray}
{\cal L}_N^{\rm int}(x)
&=&\bar N(x)\int d\xi_1\int d\xi_2 F(\xi^2_1+\xi^2_2)\\
&\times&\sum_{I=V,T}
g_N^I J_N^I (x-2\xi_1,x+\xi_1-\sqrt{3}\xi_2,x+\xi_1+\sqrt{3}\xi_2)
+{\rm h.c.}
\nonumber
\end{eqnarray}
with $J_N^V$ and $J_N^T$ being the vector and tensor currents, respectively.
The currents are symmetric under permutation of any two quarks
\begin{eqnarray}
J_N^V (y_1,y_2,y_3)&=&\vec\tau\gamma^\mu\gamma^5q^{a_1}(y_1)
q^{a_2}(y_2)\tau_2\vec\tau C\gamma_\mu q^{a_3}(y_3)\varepsilon^{a_1a_2a_3},
\nonumber\\
J_N^T (y_1,y_2,y_3)&=&\vec\tau\sigma^{\mu\nu}\gamma^5
q^{a_1}(y_1)q^{a_2}(y_2)\tau_2\vec\tau
C\sigma_{\mu\nu}q^{a_3}(y_3)\varepsilon^{a_1a_2a_3}
\nonumber
\end{eqnarray}
\noindent The notation for the spin-flavour structure of the nucleon
currents is the same as in \cite{IofSmil}. It was shown \cite{FBS,Diquark}
that the tensor current $J^T_N$ is more suitable for the
description of the data. For this reason we will use the tensor current
in the approach developed in this paper. The nucleon-three-quark coupling
$g_N^T$ is calculated from the {\it compositeness condition}
\cite{Aniv1}-\cite{PSI} which means that the
renormalization constant of the nucleon wave function is equal to zero,
$Z_N=1-\Sigma^\prime_N(m_N)=0$, with $\Sigma_N$ being the nucleon mass
operator.

The momentum distribution of the constituents in nucleons is described
by an effective relativistic vertex function $F(\xi^2_1+\xi^2_2)$.
Its shape is chosen to guarantee ultraviolet convergence.
At the same time the vertex function is a phenomenological description of
the long distance QCD interactions between quarks and gluons.
The choice of variables in the vertex function implies
the use of the center of mass frame. In this work a Gaussian vertex
function is used
\begin{eqnarray}
F(\xi^2_1+\xi^2_2)=\int\frac{d^4k_1}{(2\pi)^4}\int\frac{d^4k_2}{(2\pi)^4}
\exp(-ik_1\xi_1-ik_2\xi_2)F(k^2_1+k^2_2)
\exp\biggl(\frac{k^2_1+k^2_2}{\Lambda_N^2}\biggr)
\nonumber
\end{eqnarray}
where $\Lambda_N$ is the Gaussian range parameter which is
related to the size of the nucleon.

\begin{figure}[htbp]
\unitlength=0.4mm
\special{em:linewidth 0.4pt}
\linethickness{0.4pt}
\begin{center}
\begin{picture}(127.00,130.00)
\put(70.00,52.00){\oval(48.00,16.00)[]}
\put(45.00,52.00){\circle*{5.20}}
\put(95.00,52.00){\circle*{5.00}}
\put(95.00,52.00){\line(1,0){25.00}}
\put(120.00,53.00){\line(-1,0){25.00}}
\put(95.00,51.00){\line(1,0){25.00}}
\put(45.00,51.00){\line(-1,0){25.00}}
\put(20.00,52.00){\line(1,0){25.00}}
\put(45.00,53.00){\line(-1,0){25.00}}
\put(44.00,53.00){\line(3,5){26.00}}
\put(96.00,53.00){\line(-3,5){26.00}}
\put(70.00,96.00){\circle*{2.00}}
\put(70.00,96.00){\line(0,1){3.00}}
\put(70.00,100.00){\line(0,1){3.00}}
\put(70.00,104.00){\line(0,1){3.00}}
\put(70.00,108.00){\line(0,1){3.00}}
\put(70.00,112.00){\line(0,1){3.00}}
\put(92.00,107.00){\makebox(0,0)[cc]{q=p-p$^\prime$}}
\put(70.00,121.00){\makebox(0,0)[cc]{$\gamma$}}
\put(13.00,52.00){\makebox(0,0)[cc]{{\bf N}}}
\put(127.00,52.00){\makebox(0,0)[cc]{{\bf N}}}
\put(108.00,58.00){\makebox(0,0)[cc]{p}}
\put(32.00,58.00){\makebox(0,0)[cc]{p$^\prime$}}
\put(32.00,52.00){\line(2,1){9.00}}
\put(32.00,52.00){\line(2,-1){9.00}}
\put(108.00,52.00){\line(2,1){9.00}}
\put(108.00,52.00){\line(2,-1){9.00}}
\put(68.00,60.10){\line(4,1){9.00}}
\put(68.00,60.10){\line(4,-1){9.00}}
\put(68.00,44.10){\line(4,1){9.00}}
\put(68.00,44.10){\line(4,-1){9.00}}
\put(56.00,73.00){\line(0,2){9.00}}
\put(56.00,73.00){\line(2,1){9.00}}
\put(82.00,76.30){\line(2,-1){9.00}}
\put(82.00,76.30){\line(0,-2){9.00}}
\put(63.00,98.00){\makebox(0,0)[cc]{$\gamma_\mu$}}
\put(70.00,30.00){\makebox(0,0)[cc]{(a)}}
\end{picture}
\hspace*{3.0mm}
\begin{picture}(127.00,99.00)
\put(70.00,52.00){\oval(48.00,16.00)[]}
\put(45.00,52.00){\circle*{5.20}}
\put(95.00,52.00){\circle*{5.20}}
\put(95.00,52.00){\line(1,0){25.00}}
\put(120.00,53.00){\line(-1,0){25.00}}
\put(95.00,51.00){\line(1,0){25.00}}
\put(45.00,51.00){\line(-1,0){25.00}}
\put(20.00,52.00){\line(1,0){25.00}}
\put(45.00,53.00){\line(-1,0){25.00}}
\put(44.8,52.00){\line(0,1){3.00}}
\put(44.8,56.00){\line(0,1){3.00}}
\put(44.8,60.00){\line(0,1){3.00}}
\put(44.8,64.00){\line(0,1){3.00}}
\put(44.8,68.00){\line(0,1){3.00}}
\put(44.8,72.00){\line(0,1){3.00}}
\put(44.8,76.00){\line(0,1){3.00}}
\put(44.8,80.00){\line(0,1){3.00}}
\put(44.8,84.00){\line(0,1){3.00}}
\put(44.8,94.00){\makebox(0,0)[cc]{$\gamma$}}
\put(64.00,77.00){\makebox(0,0)[cc]{q=p-p$^\prime$}}
\put(13.00,52.00){\makebox(0,0)[cc]{{\bf N}}}
\put(127.00,52.00){\makebox(0,0)[cc]{{\bf N}}}
\put(108.00,58.00){\makebox(0,0)[cc]{p}}
\put(32.00,58.00){\makebox(0,0)[cc]{p$^\prime$}}
\put(32.00,52.05){\line(2,1){9.00}}
\put(32.00,52.05){\line(2,-1){9.00}}
\put(108.00,52.05){\line(2,1){9.00}}
\put(108.00,52.05){\line(2,-1){9.00}}
\put(70.00,50.00){\oval(50.50,35.50)[t]}
\put(68.00,67.85){\line(4,1){9.00}}
\put(68.00,67.85){\line(4,-1){9.00}}
\put(68.00,60.10){\line(4,1){9.00}}
\put(68.00,60.10){\line(4,-1){9.00}}
\put(68.00,44.10){\line(4,1){9.00}}
\put(68.00,44.10){\line(4,-1){9.00}}
\put(70.00,30.00){\makebox(0,0)[cc]{(b)}}
\end{picture}
\hspace*{3.0mm}

\vspace*{-1cm}
\begin{picture}(127.00,99.00)
\put(70.00,52.00){\oval(48.00,16.00)[]}
\put(45.00,52.00){\circle*{5.20}}
\put(95.00,52.00){\circle*{5.20}}
\put(95.00,52.00){\line(1,0){25.00}}
\put(120.00,53.00){\line(-1,0){25.00}}
\put(95.00,51.00){\line(1,0){25.00}}
\put(45.00,51.00){\line(-1,0){25.00}}
\put(20.00,52.00){\line(1,0){25.00}}
\put(45.00,53.00){\line(-1,0){25.00}}
\put(95.2,52.00){\line(0,1){3.00}}
\put(95.2,56.00){\line(0,1){3.00}}
\put(95.2,60.00){\line(0,1){3.00}}
\put(95.2,64.00){\line(0,1){3.00}}
\put(95.2,68.00){\line(0,1){3.00}}
\put(95.2,72.00){\line(0,1){3.00}}
\put(95.2,76.00){\line(0,1){3.00}}
\put(95.2,80.00){\line(0,1){3.00}}
\put(95.2,84.00){\line(0,1){3.00}}
\put(95.2,94.00){\makebox(0,0)[cc]{$\gamma$}}
\put(115.00,77.00){\makebox(0,0)[cc]{q=p-p$^\prime$}}
\put(13.00,52.00){\makebox(0,0)[cc]{{\bf N}}}
\put(127.00,52.00){\makebox(0,0)[cc]{{\bf N}}}
\put(108.00,58.00){\makebox(0,0)[cc]{p}}
\put(32.00,58.00){\makebox(0,0)[cc]{p$^\prime$}}
\put(32.00,52.05){\line(2,1){9.00}}
\put(32.00,52.05){\line(2,-1){9.00}}
\put(108.00,52.05){\line(2,1){9.00}}
\put(108.00,52.05){\line(2,-1){9.00}}
\put(70.00,50.00){\oval(50.50,35.50)[t]}
\put(68.00,67.85){\line(4,1){9.00}}
\put(68.00,67.85){\line(4,-1){9.00}}
\put(68.00,60.10){\line(4,1){9.00}}
\put(68.00,60.10){\line(4,-1){9.00}}
\put(68.00,44.10){\line(4,1){9.00}}
\put(68.00,44.10){\line(4,-1){9.00}}
\put(70.00,30.00){\makebox(0,0)[cc]{(c)}}
\end{picture}

\vspace*{-1cm}
{\bf Fig.1} Nucleon Form Factor. Triangle diagram (a), bubble
diagrams (b) and (c).
\end{center}
\end{figure}
The interaction of quarks with the electromagnetic field is introduced
by the standard minimal substitution in the free Lagrangian. The gauging
of nonlocal interactions is done by using the time-ordering P-exponent
in the (1). See details in refs. \cite{Aniv1}-\cite{PSI}. It leads to the
modification of the interaction Lagrangian which generates nonlocal vertices
which couple nucleons, photons and quarks. Therefore, the electromagnetic
form factors of nucleons are described by the standard triangle diagram
(Fig.1a) and, additionally, by the bubble or contact diagrams
(Fig.1b and Fig.1c).

We choose the standard form of quark propagator
$S_q(k)=(m_q-\not\! k)^{-1}$,
corresponding to a free fermionic field with mass $m_q$.
For the time being we shall
avoid the appearance of unphysical imaginary parts in the Feynman diagrams
by restricting the calculations to $m_N<3m_q$.
Thus, there are two adjustable parameters in our model: the constituent
quark mass $m_q$ and the range parameter $\Lambda_N$. We shall determine
them by fitting the electromagnetic properties of nucleons.

\section{Numerical Results and Discussion}

Within the relativistic quark model described above we shall evaluate
the magnetic moments, the charge radii
and the behavior of the electric and magnetic form factors for
$0\le Q^2\le 1$ GeV$^2$.
It is convenient to separate the contributions from the triangle diagram
(Fig.1a) and the bubble diagrams (Fig.1b and 1c) as follows
\begin{eqnarray}
\Lambda_\mu(p,p^\prime)=
\frac{q_\mu}{q^2}[\Sigma_N (p)-\Sigma_N (p^\prime)]+
\Lambda^\perp_{\mu,\Delta}(p,p^\prime)+
\Lambda^\perp_{\mu, {\rm bubble}}(p,p^\prime)
\end{eqnarray}
\noindent The functions $\Lambda^\perp_{\mu, \Delta}(p,p^\prime)$ and
$\Lambda^\perp_{\mu, {\rm bubble}}(p,p^\prime)$
are the modified triangle and bubble vertex functions which are
orthogonal to the photon momentum $q^\mu \Lambda^\perp_\mu(p,p^\prime)=0$.
We shall work in the limit of isospin
invariance where the masses of $u$ and $d$ quarks are equal.
In a first step we set the parameters of our model
$m_q$ and $\Lambda_N$ to the well measured
static properties of the nucleons, to the magnetic
moments and to the charge radii \cite{PDG,Dumb}. It was found
that the best fit of data is obtained for $m_q=$420 MeV and
$\Lambda_N$=1.25 GeV. In Table 1 the best fit of the nucleon static
properties is compared to other theoretical approaches.

\newpage
\begin{center}
{\bf Table 1. Static Properties Compared to Theoretical Approaches}
\end{center}
\def\arraystretch{1.2}
\begin{center}
\begin{tabular}{|c|c|c|c|c|c|c|}
\hline\hline
Approach & $\mu_p$ & $\mu_n$ & $r^p_E$, fm & $<r^2>^n_E$, fm$^2$ &
$r^p_M$, fm & $r^n_M$, fm \\
\hline\hline
 Our & 2.79 & -1.86 & 0.92 & -0.132 & 0.84 & 0.84 \\
\hline
 Ref. \cite{IofSmil} & 2.96 & -1.93 & & & & \\
\hline
 Ref. \cite{Balit} & 2.80 & -1.95 & & & & \\
\hline
 Ref. \cite{Meis} & 2.77 & -1.84 & 0.97 & -0.25 & 0.94 & 0.94 \\
\hline
 Ref. \cite{Azna} & 2.811 & -1.848 & 0.811 & -0.094 & 0.825 & 0.781 \\
\hline
 Ref. \cite{Baric} & 2.73 & -1.975 & 0.85 &  &  & \\
\hline\hline
 Exp. & 2.79 & -1.91 & 0.86$\pm$0.01 & -0.119$\pm$0.004 & 0.86$\pm$0.06
      & 0.88$\pm$0.07 \\
\hline\hline
\end{tabular}
\end{center}

In a second step we calculate the nucleon electromagnetic form factors for
the range $Q^2\le $1 GeV$^2$.
The shape of the electromagnetic nucleon form
factors in the region $0\le Q^2\le$ 1 GeV$^2$ for
the same optimal values of the parameters,
$m_q$=420 MeV and $\Lambda_N$=1.25 GeV, is shown in Figs.2-5.
The normalized magnetic form factors of nucleons $G_M^N(Q^2)/G_M^N(0)$
are plotted in Fig.2 and Fig.3.

\begin{figure}[htbp]
\begin{center}
\vspace*{-45mm}
\mbox{\epsfysize=11cm\epsffile{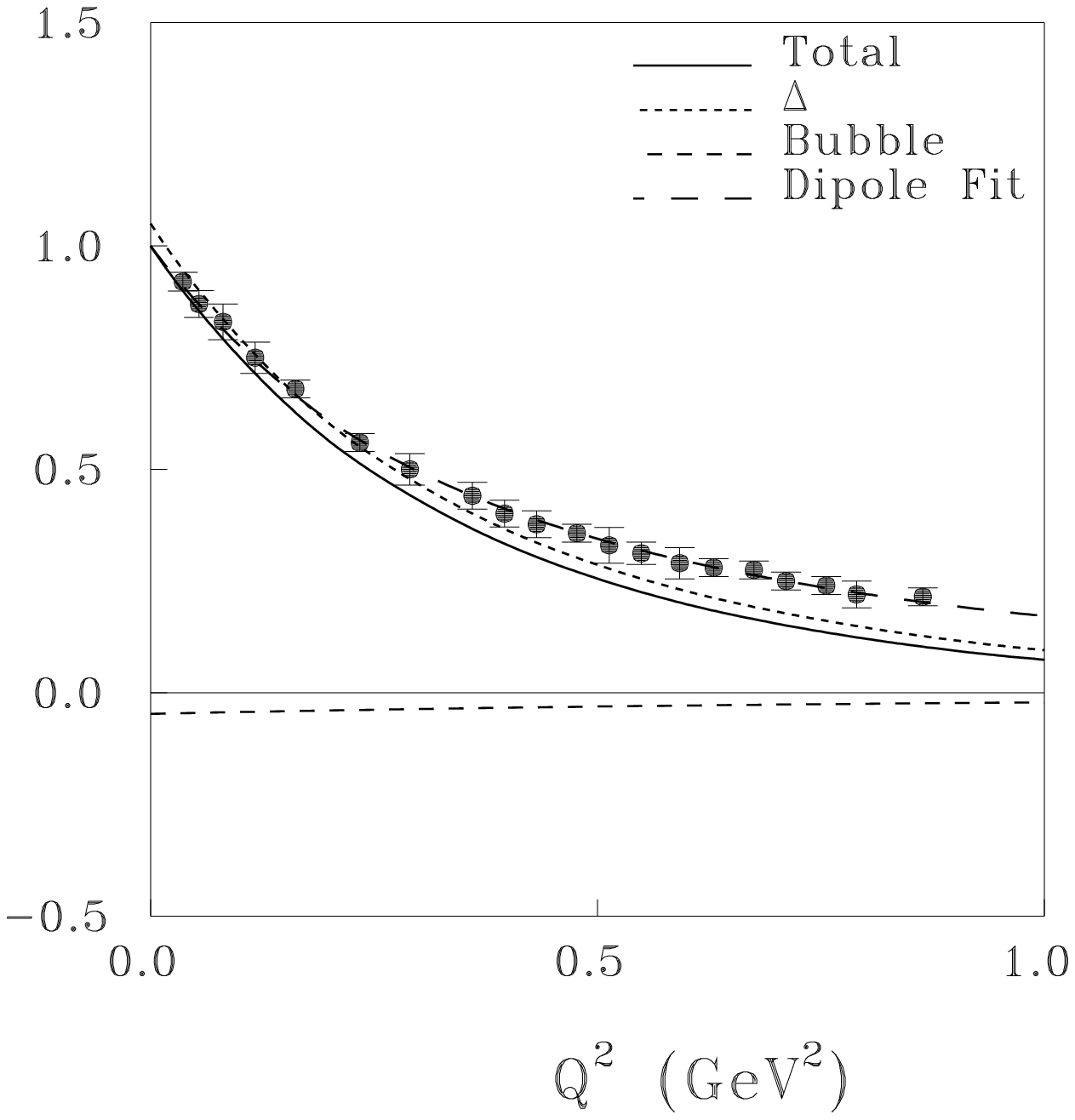}}
\hspace*{-10mm}
\mbox{\epsfysize=11cm\epsffile{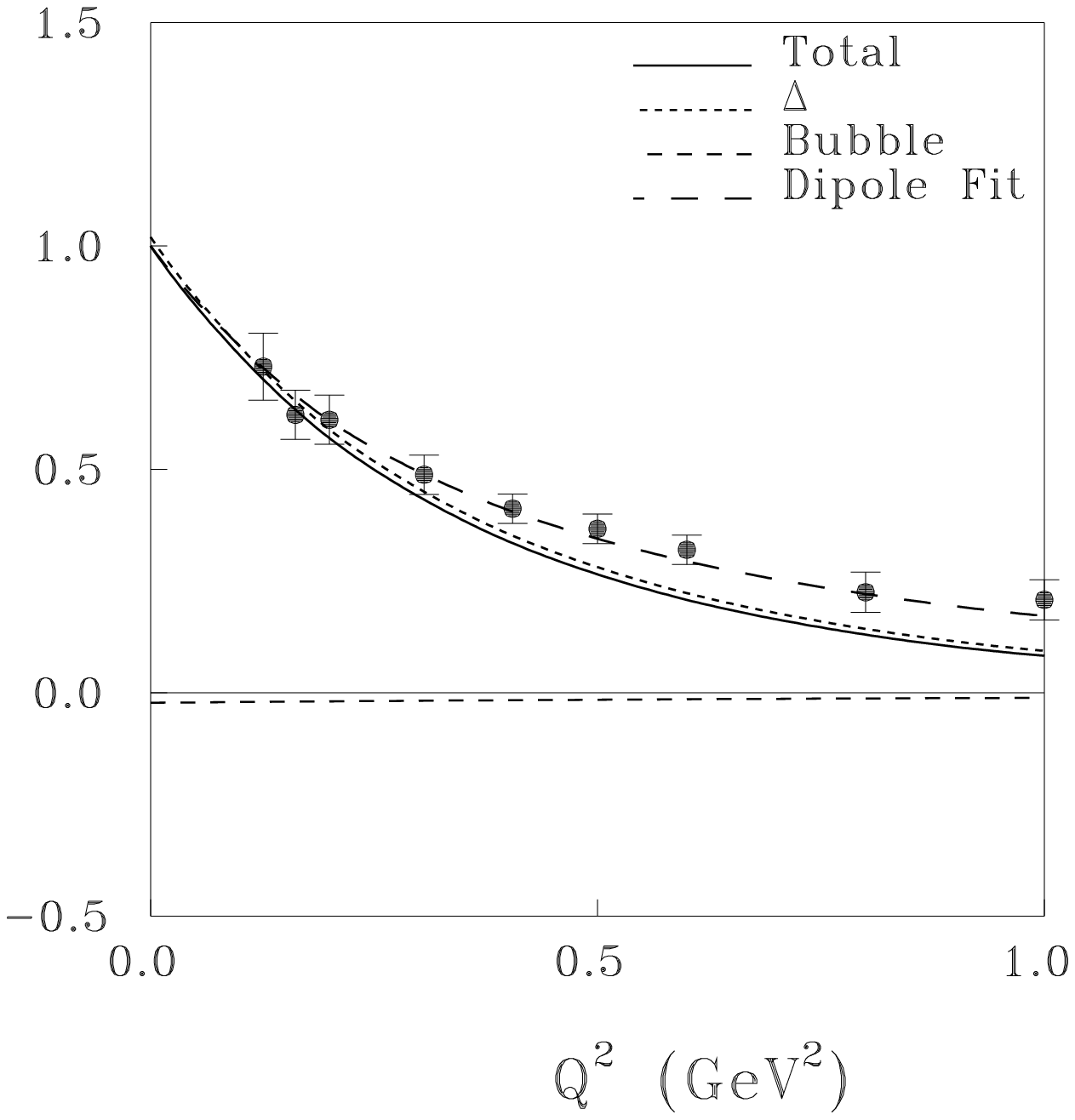}}
\\
\vspace*{7mm}
\mbox{\hspace*{.8cm}{\bf Fig.2}. Magnetic form factor of proton.}
\hspace*{1cm}
\mbox{{\bf Fig.3}. Magnetic form factor of neutron.}

\mbox{\epsfysize=11cm\epsffile{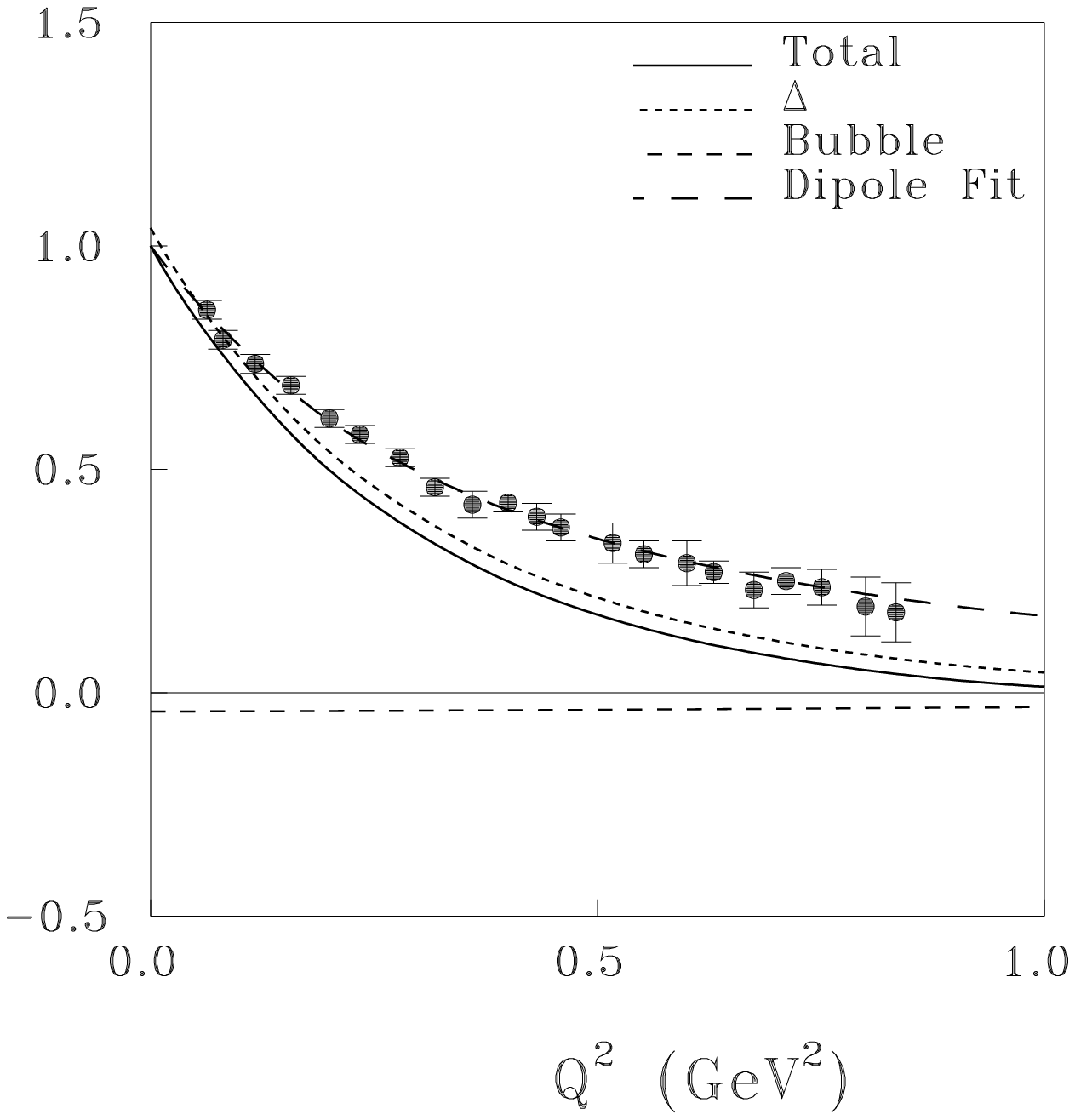}}
\hspace*{-10mm}
\mbox{\epsfysize=11cm\epsffile{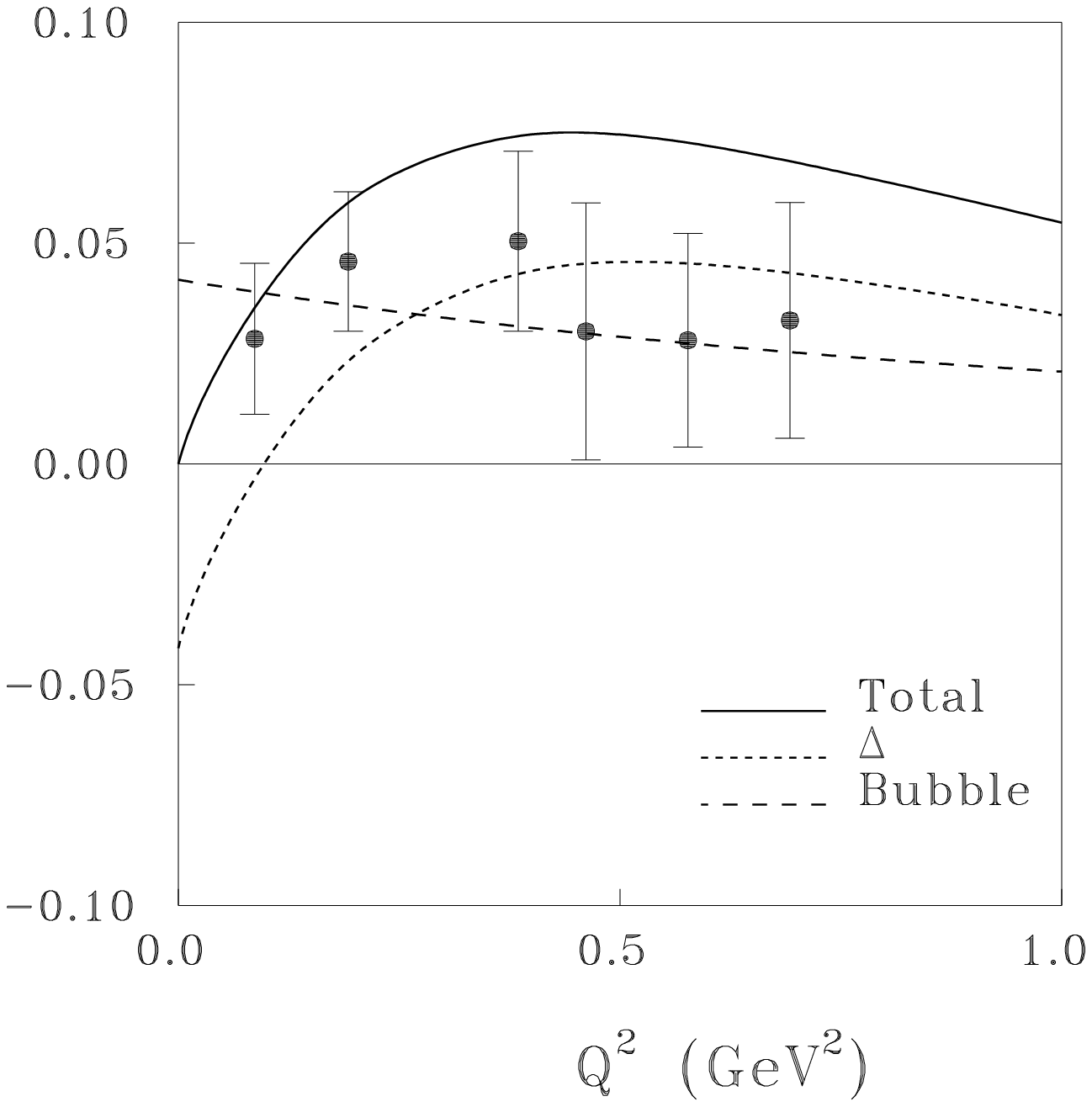}}
\\
\vspace*{7mm}
\mbox{\hspace*{.8cm}{\bf Fig.4}. Electric form factor of proton.}
\hspace*{1cm}
\mbox{{\bf Fig.5}. Electric form factor of neutron.}
\end{center}
\end{figure}

The short dashed line corresponds to the contribution of the
triangle vertex function $\Lambda^\perp_{\mu, \Delta}(p,p^\prime)$
whereas the medium-long dashed line gives the contribution of the bubble
or contact term $\Lambda^\perp_{\mu, {bubble}}(p,p^\prime)$.
This contribution is roughly constant and not large.
It matters however for the details of the fitting procedure.
The total contribution is marked by the solid line.
The long-dashed line is the dipole approximation
$D(Q^2)=[1+Q^2/0.71\mbox{GeV}^2]^{-2}$ which fits the data taken from
\cite{DATA} quite well.

The electric form factors of the proton and the neutron are shown
in Fig.4 and 5, respectively.
It is seen that the electric proton and magnetic nucleon form factors
fall faster than the dipole fit of the experimental data
for momentum transfers $0\le Q^2\le 1$ GeV$^2$.
(The behavior of our form factors scaled by the dipole fit is shown
in Fig.6).

\begin{figure}[htbp]
\begin{center}
\vspace*{-45mm}
\mbox{\epsfysize=12.cm\epsffile{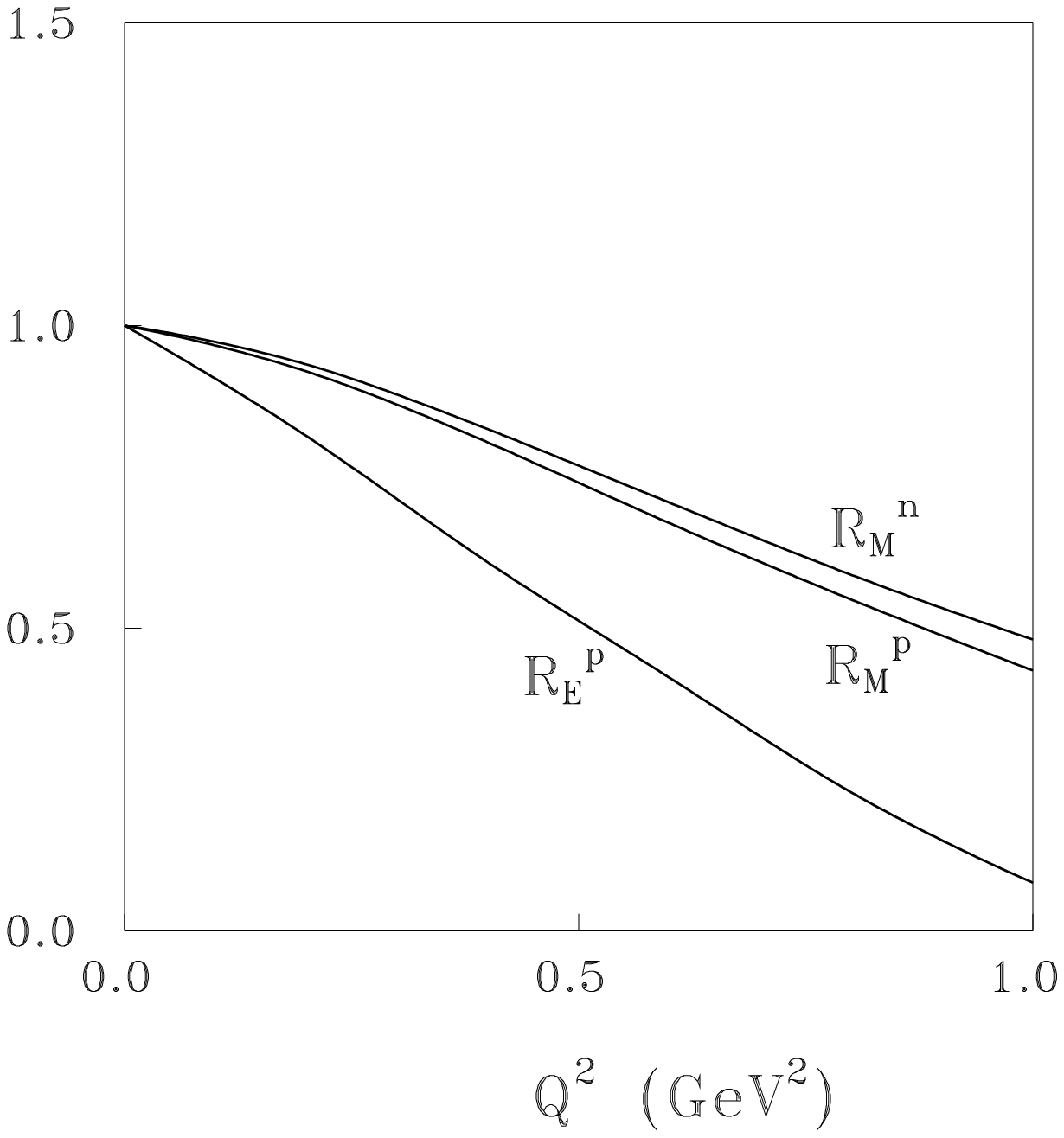}}
\\
\vspace*{7mm}
\mbox{{\bf Fig.6}. Nucleon form factors
scaled by the dipole fit:}\\
$R_E^p$ - \mbox{electric form factor of proton,}\\
$R_M^p$ - \mbox{magnetic form factor of proton,}\\
$R_M^n$ - \mbox{magnetic form factor of neutron}
\end{center}
\end{figure}

First, one has to remark that the Dirac and Pauli form factors have similar
behavior in an approach based on calculation of the Feynman quark digram like
in Fig. 1 disregarding to the choice of the vertex function because of
they are defined by the $Q^2$-dependence of the vector and tensor functions,
respectively. It gives that the electric proton form factor goes
to zero at $Q^2\approx 1.5$ GeV$^2$.
Second, both the Dirac and Pauli form factors are turned out to fall faster
than the dipole approximation. This may be seen as the result either of using
the Gaussan shape for vertex function used here or neglecting
the contribution of the pionic cloud.

\section{Conclusion}

The electromagnetic form factors of nucleons have been calculated
within a relativistic three quark model with Gaussian shape
for the nucleon-quark vertex, and standard (non-confined) quark propagators.
Gauge invariance of the nonlocal hadron-quark interaction has been
implemented by the path-independent definition for
the derivative of the time-ordering P-exponent.
The allowed region for the two adjustable parameters,
the range parameter $\Lambda_N$ appearing in the Gaussian and
the constituent quark mass $m_q$, has been obtained by fitting
the data for the magnetic moments and the electromagnetic radii of
the nucleons.
It is found that their values calculated with $m_q$=420 MeV and
$\Lambda_N$=1.25 GeV agree very well with the experimental data.

It is turned out that the electric proton and magnetic nucleon form factors
fall faster than the dipole fit of the experimental data
for momentum transfers $0\le Q^2\le 1$ GeV$^2$.
This may be seen as the result either of using
the Gaussan shape for vertex function used here or neglecting
the contribution of the pionic cloud.

\vspace*{1.5cm}
\section{Acknowledgements}

We would like to thank P. Kroll, R. Rosenfelder, J. Bolz, H. Kamada
and H.-Q. Zheng for useful discussions.
This work was supported in part by the INTAS Grant 94-739,
by the Russian Fund of Fundamental Research (RFFR)  under
contracts 94-02-03463-a, 96-02-17435-a and
State Committee of Russian Federation for
Education (project N 95-0-6.3-67,
Grand Center at S.-Petersburg State University).

\end{document}